\documentclass[sigconf]{acmart}

\usepackage[noabbrev]{cleveref}

\usepackage{tabularray}
\UseTblrLibrary{booktabs}
\usepackage{multirow}

\AtBeginDocument{%
  }

\copyrightyear{2026}
\acmYear{2026}
\setcopyright{cc}
\setcctype{by}
\acmConference[SIGCSE TS 2026]{Proceedings of the 57th ACM Technical Symposium on Computer Science Education V.1}{February 18--21, 2026}{St. Louis, MO, USA}
\acmBooktitle{Proceedings of the 57th ACM Technical Symposium on Computer Science Education V.1 (SIGCSE TS 2026), February 18--21, 2026, St. Louis, MO, USA}
\acmPrice{}
\acmDOI{10.1145/3770762.3772633}
\acmISBN{979-8-4007-2256-1/26/02}

\begin{document}

\title[Complexity of Code Structuring Exercises]{Systematically Thinking about the Complexity of Code Structuring Exercises at Introductory Level}



\author{Georgiana Haldeman}
\orcid{0000-0001-6046-5924}
\affiliation{%
  \institution{Colgate University}
  \city{Hamilton, NY}
  \country{USA}}
\email{ghaldeman@colgate.edu}

\author{Peter Ohmann}
\orcid{0000-0002-7670-7374}
\affiliation{%
  \institution{College of St. Benedict / St. John's University}
  \city{St. Joseph, MN}
  \country{USA}}
\email{pohmann001@csbsju.edu}

\author{Paul Denny}
\orcid{0000-0002-5150-9806}
\affiliation{%
  \institution{University of Auckland}
  \city{Auckland}
  \country{New Zealand}}
\email{paul@cs.auckland.ac.nz}


\begin{abstract}
  Decomposition and abstraction (DA) is an essential component of computational thinking, yet it is not always emphasized in introductory programming courses.  In addition, as generative AI further reduces the focus on syntax and increases the importance of higher-level code reasoning, there is renewed opportunity to teach DA explicitly. In this paper, we introduce a framework for systematically assessing the complexity of code structuring tasks, where students must identify and separate meaningful abstractions within existing, unstructured code. The framework defines three dimensions of task complexity, each with multiple levels: repetition, code pattern, and data dependency. To support practical use, we provide example tasks mapped to these levels and offer an interactive tool for generating and exploring DA problems.  The framework is designed to support the development of educational tasks that build students' skills with DA in the procedural paradigm.

\end{abstract}




\maketitle

\section{Introduction}

Decomposition and abstraction (DA) is central to computational thinking~\cite{wing_computational_2006}. It spans across broad areas, such as: (1) software development (dividing a software project into modules, functions, or classes), (2) algorithm or data structure design (breaking down an algorithm into simpler steps or subroutines, or a data structure into simpler parts), and (3) problem-solving in general (breaking down a real-world problem into smaller, more manageable tasks).

While often 
deferred until later courses,
DA has become a central topic in newer curricula of courses at the introductory level ~\cite{chiodini_teaching_2023} and in particular those that integrate generative AI~\cite{vadaparty_cs1-llm_2024}. These newer curricula have a reduced emphasis on syntax and writing code, and an increased emphasis on skills needed to successfully produce software with an LLM, such as reading and explaining code, testing code, and decomposing large problems into small functions that are solvable by an LLM ~\cite{franklin2025generative, vadaparty_cs1-llm_2024}.

However, these proposed curricula ~\cite{vadaparty_cs1-llm_2024} and associated CS1 textbooks ~\cite{porter2024learn} focus primarily on the top-down approach to decomposition and we believe that achieving competency in both top-down and bottom-up approaches is needed. Studies have shown that expert programmers use both types of strategies; in particular, they use a top-down approach whenever they have a suitable schema / plan available (e.g., recalling a very similar structure from previously completed projects) and a bottom-up approach in the absence of suitable schemata (e.g. for unfamiliar or particularly difficult problems) ~\cite{jorgensen_top-down_2004,robins201912}. Robins concludes that \textit{programmers can use a mixture of these strategies as they work on familiar or unfamiliar subproblems}~\cite{robins201912}.

\begin{table*}
\caption{Sources considered in the reflection on what makes a good code abstraction for procedural code at introductory level.}
\label{tab:sources}
\centering
\begin{tblr}{rowsep=.5pt,
  colspec={p{.05\linewidth}p{.35\linewidth}p{.35\linewidth}p{.13\linewidth}},
  row{1}={font=\bfseries},
  column{1}={font=\bfseries},
  row{2,3,4,5,7,9,11,13}={bg=gray!10},
}
\toprule
\textbf{Source} & \textbf{Paper/ Book} & \textbf{Foci}   & \textbf{Type of Pattern} \\ 
\midrule
1 &  \SetCell[r=4]{l}{{{Iyer and Zilles 2021 Pattern Census: A Characterization of Pattern Usage in Early Programming Courses \cite{iyer_pattern_2021}}}} & 
\url{https://users.cs.duke.edu/~ola/patterns/plopd/loops.html} 
& Programming              \\ 
2 & & 
\url{https://csis.pace.edu/~bergin/patterns/Patternsv5.html}  
& Programming              \\ 
3 &  & \url{https://saja.kapsi.fi/var\_roles/role\_intro.html}            & Roles of Variables       \\ 
4 &   & Table 2 & Mixed  \\ 
5 & Xie et al. 2019 A theory of instruction for introductory programming skills \cite{xie_theory_2019} & Section 4.2 & Algorithmic \\ 
6 & Harrer et al. 2018 Java By Comparison: Become a Java Craftsman in 70 Examples \cite{harrer2018java} & pages 7-9 and 114-117                                                                      & Refactoring              \\ 
7 & Ginat et al. 2013 Novice Difficulties with Interleaved Pattern Composition         \cite{ginat2013novice}                         & Section 2                                                                                  & Algorithmic              \\ 
8                                     & Muller et al. 2007 Pattern-Oriented Instruction and its Influence on Problem Decomposition and Solution Construction      \cite{muller2007pattern}                           & Section 3                                                                                  & Algorithmic              \\ 
9                                     & Muller 2005 Pattern Oriented Instruction and the Enhancement of Analogical Reasoning  \cite{muller_pattern_2005}                                      & Table 1                                                                                    & Algorithmic              \\ 
10                                    &  McConnell 2004 Code Complete: A Practical Handbook of Software Construction  \cite{mcconnell2004code}                                    & Chapters 7 and 24                                                                          & NA                       \\ 
11                                    & Fowler 2018 Refactoring: Improving the Design of Existing Code \cite{fowler2018refactoring} & Chapters 1-3 and 6                                                                         & NA                       \\ 
12                                    & Martin 2008 Clean Code: A Handbook of Agile Software Craftsmanship \cite{martin2009clean}                                        & Chapter 3                                                                                  & NA                       \\ 
\bottomrule
\end{tblr}
\end{table*}


In our prior work, we proposed a complementary conceptual framework for decomposition that employs the bottom-up approach ~\cite{haldeman2025teaching}. For simplicity, we refer to bottom-up decomposition as \textit{code structuring} in the rest of the paper. This work includes an educational resource with scaffolded exercises that target code structuring skills. However, what is currently missing is a clear classification scheme that defines what makes such exercises more or less complex to solve.  Such a scheme would allow educators to design and sequence tasks to match the learning needs of their students. In this paper, we seek to fill this gap by proposing a framework for classifying the complexity of code structuring tasks at the introductory level. 


We believe that code structuring is an essential skill for software development. In software development, it appears in software design (which is predominantly a top-down approach) or refactoring (which is predominantly a bottom-up approach). Code refactoring is the process of restructuring existing source code (or changing the factoring) without changing its external behavior ~\cite{mcconnell2004code}. In addition, we know that many introductory students struggle with the structuring of their code from both studies ~\cite{izu2025introducing} and anecdotally. 
Even when students can write functionally correct programs, they often produce code that lacks readability and structure, and targeted refactoring resources have been shown to help \cite{izu2022resource}.
This struggle does not improve later in advanced courses. Shah et al.~\cite{shah2025identifying} found that two-thirds of the students in an upper-division software engineering course had long functions in their projects.


Lastly, a systematic approach for the generation of code structuring exercises presents many benefits: (1) it provides students with 
hands-on
practice of a targeted skill, namely, identifying and extracting new abstractions from code; (2) it enables the creation of new and isomorphic problems, rubrics for grading, and feedback generation; and (3) the assessments de-emphasize syntax and therefore have a lower cognitive load similar to Parson's problems ~\cite{ericson2022parsons}. While Parson's problems ~\cite{ericson2022parsons} primarily target the ordering of code snippets, the assessments targeted by our framework require students to isolate and abstract code snippets into separate methods.

The remainder of this paper is organized as follows. \Cref{sec:relatedwork} discusses background and related work. \Cref{s:reflection} makes a case for what makes a good procedural code abstraction at the introductory level.  \Cref{sec:framework} describes the proposed framework and provides examples of entry-level problems that span our pattern dimensions. \Cref{s:usage} explains how the framework and the companion interactive repository can be used by the community. \Cref{s:future} details future directions of this work. \Cref{sec:conclusion} provides an overview of the paper and concludes.

\section{Related Work}
\label{sec:relatedwork}

Decomposition and abstraction (DA) appears in many areas of computer science and Computing Education Research (CER). The closest areas to our work are: code writing (in particular, algorithmic patterns and roles of variables) and code quality (in particular, refactoring and code properties, such as code cohesion).

\subsection{Code writing}

Decomposition is at the core of code writing, as emphasized by Pattern-Oriented Instruction (POI) ~\cite{muller_almost_2004,muller_pattern_2005,muller2007pattern} and the theory of instruction for introductory programming skills ~\cite{xie_theory_2019}. Patterns can be leveraged in problem decomposition (the top-down approach) or the structuring of code (the bottom-up approach).  Several types of patterns have been identified in the literature. A summary of 
publications that specify patterns or provide guidelines is provided in Table ~\ref{tab:sources}.

\subsubsection{Pattern-Oriented Instruction and Algorithmic Patterns}

POI is a pedagogical approach that refers to a methodology by which algorithmic patterns are incorporated into a course's instruction ~\cite{muller2007pattern}. \textit{Algorithmic patterns} represent good examples of elegant and efficient solutions to recurring algorithmic problems. Muller et al. ~\cite{muller2007pattern} have shown that the incorporation of algorithmic patterns through POI enhances the construction of algorithmic problem-solving knowledge.



\subsubsection{Other Programming Patterns and Roles of Variables}

In parallel with the research on POI, others have independently recognized the value of patterns for the development of essential thinking skills in programming. Proulx ~\cite{proulx_programming_2000} was the first to introduce a collection of programming and design patterns to be used in the teaching of programming. Sajaniemi~\cite{sajaniemi_roles_2005} studied patterns that are focused on the specific role of variables and their impact on learning to program. Finally, Iyer and Zilles~\cite{iyer_pattern_2021} compiled a repository of patterns at the introductory level and found significant agreement among the studied courses about which patterns should be learned, and  that most of this consensus is language independent.


\subsection{Code quality}


Decomposition is also discussed under the topic of code quality in software engineering.
Several books on code quality, and software engineering more broadly, provide guidelines and principles for structuring code. Code Complete~\cite{mcconnell2004code}, Clean Code~\cite{martin2009clean} and Refactoring~\cite{fowler2018refactoring} are three of the most established software engineering books with tens of thousands of citations collectively. Java by Comparison ~\cite{harrer2018java} is a book on refactoring patterns similar to Refactoring~\cite{fowler2018refactoring} aimed at introductory courses.

When it comes to measures of well-structured code, code cohesion is high on the list. Introduced in 1974 by Stevens, Myers and Constatine~\cite{stevens1974structured}, the idea of cohesion with its different levels has stood the test of time, being mentioned in notable books such as McConnell's \textit{Code Complete: A Practical Handbook of Software Construction} in 2004~\cite{mcconnell2004code} and the latest edition of \textit{Computer Science: A structured Programming Approach in C} by Afyouni and Forouzan~\cite{forouzan1999computer} published in 2023. McConnell~\cite{mcconnell2004code} claims: \textit{...cohesion is still alive and well as the workhorse design heuristic at the individual-routine level. For routines, cohesion refers to how closely the operations in a routine are related.} There are many levels of code cohesion out of which functional cohesion is \textit{the strongest and best kind of cohesion, occurring when a routine performs one and only one operation and moreover what their names say they do} ~\cite{mcconnell2004code}.

Novices may benefit from learning about code quality, including potentially increasing the efficiency of knowledge transfer and learning~\cite{katona_clean_2021, izu2025introducing}. Many novices and advanced students write code with poor style ~\cite{izu2025introducing,shah2025identifying}. Katona ~\cite{katona_clean_2021} found that students made fewer code comprehension mistakes when reading a properly styled code base and observed that \textit{significantly less ... 
processing 
[was] necessary to understand the code.} However, Tempero et al.~\cite{tempero2024comprehensibility} found that the influence of function decomposition on code understanding is inconclusive, suggesting that functional decomposition does not universally enhance code comprehensibility.




\begin{figure}[h]
    \centering
    \includegraphics[width=.5\textwidth]{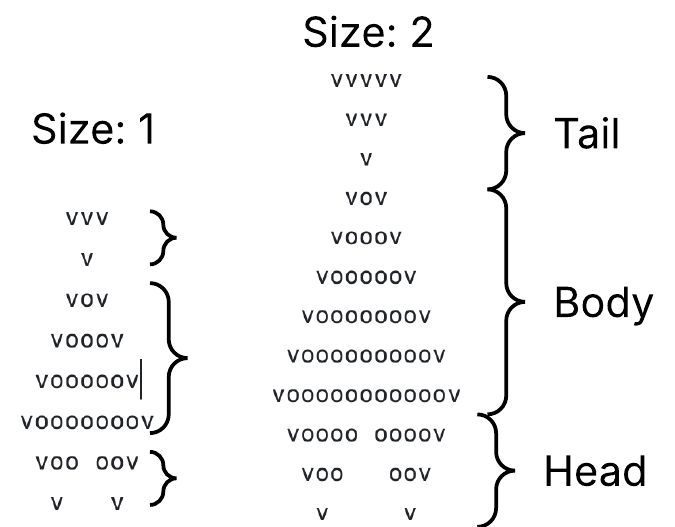}
    \caption{Example of a decomposition task with repetition that is scaled by a variable amount (Level 3 on the Repetition Dimension in Figure~\ref{fig:repetition}). For this assignment, the goal is to arrive at a decomposed program that can draw fish of different sizes using ASCII characters. The program can be first decomposed into three functions that draw different parts of the fish. Next, note that each part is scaled vertically and horizontally in relation to the size of the fish. More specifically, the tail and the head are one row larger than the size and the body is twice that vertically. Horizontally, the width of the fish is $4*(size+1)+3$ with increasing/decreasing patterns of two characters between consecutive rows.}
    \label{fig:fish}
\end{figure}

\begin{figure*}[h]
    \centering
    \begin{tabular}{@{}c@{}}
        \includegraphics[height=.22\linewidth]{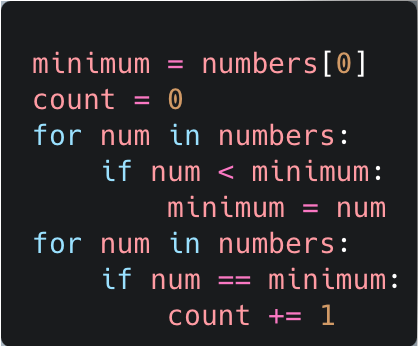} \\[\abovecaptionskip]
        \small (a) Concatenation Pattern Composition
    \end{tabular}
    \begin{tabular}{@{}c@{}}
        \includegraphics[height=.22\linewidth]{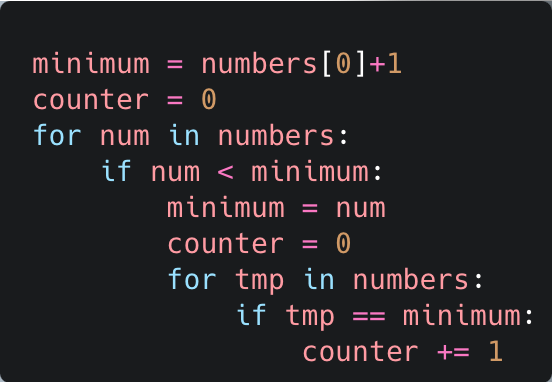} \\[\abovecaptionskip]
        \small (b) Inclusion Pattern Composition
    \end{tabular}
    \begin{tabular}{@{}c@{}}
        \includegraphics[height=.22\linewidth]{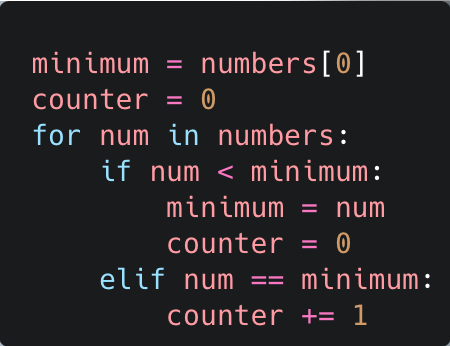} \\[\abovecaptionskip]
        \small (c) Interleaved Pattern Composition
    \end{tabular}
    \caption{Code examples of the three pattern composition types: concatenation, inclusion and interleaved for the problem of counting the occurrences of the minimum value in a list of numbers. There are two patterns for this problem: finding the minimum and counting how many times it appears in the list. The solution can compose the two patterns in different ways: first the minimum pattern then the counting pattern (concatenation), the counting pattern gets executed every time a new minimum is found (inclusion), or swap between the two patterns whenever a new minimum has been found (interleaved).} 
    \label{tab:pattern_composition}
\end{figure*}

\section{What makes a good procedural abstraction?}
\label{s:reflection}


Structuring code encompasses reasoning about abstractions. In other words: What makes a piece of code a good candidate to be made into a function? To answer this question, we looked into ways to structure and refactor code, namely, we looked at established CER papers on patterns, Pattern-Oriented Instruction, roles of variables, and software engineering books (refer to Section~\ref{sec:relatedwork} for more details.) Table~\ref{tab:sources} indicates which papers and books we considered most closely when conceptualizing our framework. We also considered other software engineering books, such as, \textit{Design Patterns: Elements of Reusable Object Oriented Software}~\cite{gamma1995design}, \textit{Refactoring to Patterns}~\cite{kerievsky2005refactoring}, and \textit{Implementation Patterns} ~\cite{beck2007implementation}. However, we concluded that these books are primarily focused on design and practical guidelines and less so on structuring code in the procedural paradigm, which is the scope of this paper.


In software engineering, one primary goal of structuring code is to redistribute complexity. Code complexity can be horizontal or vertical. Horizontal complexity is when code statements combine complex expressions. Vertical complexity is related to function size. Code complexity is often assessed using McCabe's cyclomatic complexity~\cite{mccabe1976complexity} or lines of code (LOC) ~\cite{yan_closer_2023}. In this work, we primarily focus on vertical complexity, which is reflected by LOC. This limitation is negligent given that any piece of code with horizontal complexity can be re-written into a version with vertical complexity and the number of refactoring patterns involving horizontal complexity are far fewer (for example, simplifying a boolean expression, pages 7-9 in ~\cite{harrer2018java}).

Regarding what makes a good procedural abstraction, Robert C. Martin writes in \textit{Clean Code}~\cite{martin2009clean}: 

\begin{quote}
\textit{FUNCTIONS SHOULD DO ONE THING. THEY SHOULD DO IT WELL. THEY SHOULD DO IT ONLY...}
\end{quote}
\begin{quote}
\textit{In order to make sure our functions are doing “one thing,” we need to make sure that the statements within our function are all at the same level of abstraction. We want the code to read like a top-down narrative. We want every function to be followed by those at the next level of abstraction so that we can read the program, descending one level of abstraction at a time as we read down the list of functions. I call this "The Step-down Rule". To say this differently, we want to be able to read the program as though it were a set of TO paragraphs, each of which is describing the current level of abstraction and referencing subsequent TO paragraphs at the next level down. }
\end{quote}


This guideline is consistent with the definition of functional cohesion ~\cite{stevens1974structured, mcconnell2004code}. We reflected on these guidelines in relation to patterns. The specific question we aimed to answer was: 

\textit{Which types of patterns best fit the software engineering guidelines of functional decomposition exposed above?}

We concluded that the algorithmic patterns are the best candidates to be separated as functions because of their functional cohesion. Algorithmic patterns form building blocks for developing algorithms~\cite{muller_pattern_2005} similarly to how functions form building blocks for programs. One example of an algorithmic pattern is \textit{Exists?} which \textit{checks for the existence of an item that satisfies a condition}~\cite{muller2007pattern}. Separating its corresponding code into a separate function seems reasonable, regardless of the problem in which it appears.

Conversely, many of the other identified patterns either do not have functional cohesion or do not align in scope with our work. For example, the \textit{Name Use} programming pattern from ~\citet{proulx_programming_2000}\textemdash{}\textit{identifiers follows a set pattern: declare - define/build - use - destroy}\textemdash{}is beyond the scope of this work because it does not involve any aspect of structuring code into procedural abstractions. Moreover, the only refactoring pattern that overlaps with our work is the \textit{Extract Function} (page 106 in ~\cite{fowler2018refactoring}).

\section{Code Structuring Complexity Framework}
\label{sec:framework}

Building on the conceptual framework proposed in our previous work~\cite{haldeman2025teaching} and using our reflections in Section ~\ref{s:reflection}, we present a theoretical framework for assessing the complexity of code structuring tasks. This framework provides a practical approach to thinking about code structuring and the aspects that make it challenging. We posit that the complexity of code structuring tasks varies along three dimensions: repetition, code patterns, and data dependency. We label each level of each dimension from 0 to 3 to signify increased complexity without making any claims about the relative increase from one level to another or the relative comparison between dimensions. In other words, we are not implying that a level one in one dimension is equivalent to the level one in the other dimensions or that the increase in complexity from level zero to level one is the same as the increase in complexity from level two to three.

\begin{figure}[H]
    \centering
    \includegraphics[width=.9\columnwidth]{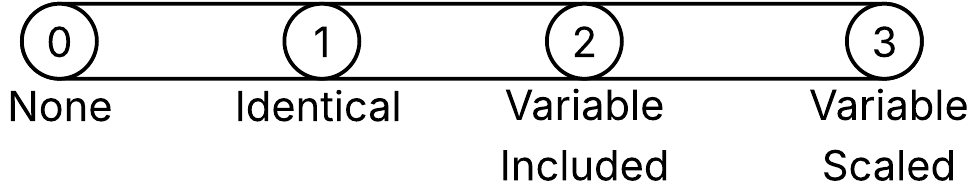}
    \caption{Levels of the Repetition Dimension. Identical Level 1 is represented by tasks in which snippets of code are identical copies of each other. Variable Included Level 2 is represented by tasks in which snippets of code are near identical copies of each other except certain parts that can be passed using a parameter. Variable Scaled Level 3 is represented by tasks that are similar in structure with a linear relation between them. An example of a task at the Variable Scaled Level 3 is shown in Figure~\ref{fig:fish}.}
    \label{fig:repetition}
\end{figure}

\subsection{Repetition Dimension}

We reflect that the repetition dimension (also referenced as a criterion for code decomposition in Haldeman et al.~\cite{haldeman2025teaching}) has four levels as shown in Figure~\ref{fig:repetition}. An example of a level 3 task is shown in Figure~\ref{fig:fish}. An example of level 2 would be a program that prints the lyrics to a nursery rhyme. For example, the Old MacDonald's song repeats three times, except each time it refers to a different animal and the sounds it makes. The animal's name and sound can be passed as a parameter to an extracted function.


\begin{figure*}[h]
    \centering
    \begin{tabular}{@{}c@{}}
        \includegraphics[width=.47\linewidth]{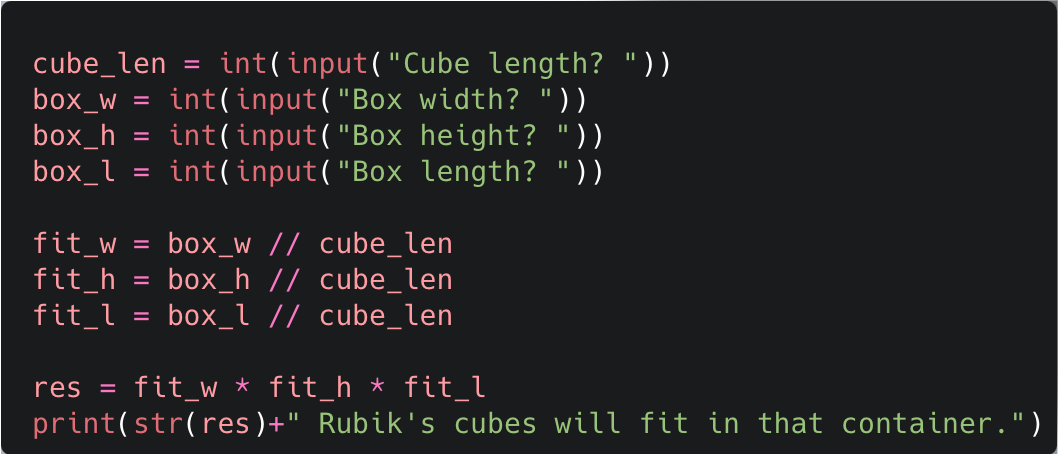} \\[\abovecaptionskip]
        \small (a) Non-decomposed Code for the Rubik's Cube Program
    \end{tabular}
    \begin{tabular}{@{}c@{}}
        \includegraphics[width=.47\linewidth]{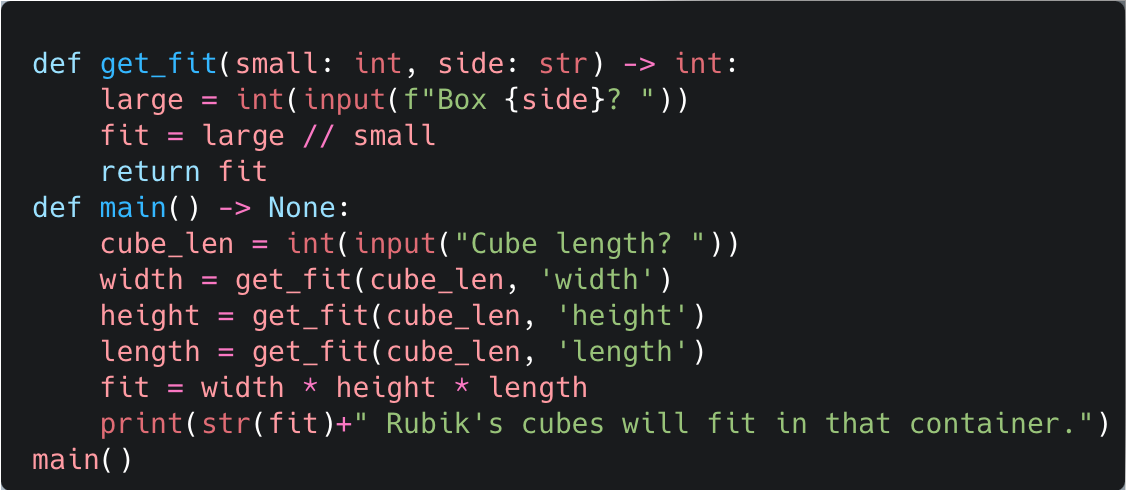} \\[\abovecaptionskip]
        \small (b) Decomposed Code for the Rubik's Cube Program
    \end{tabular}
    \caption{Code example for the Non-sequential Data Dependency dimension level. The Rubik's Cube Program  computes how many cubes can fit along all the three dimensions of a container and then multiplies them to find the number of cubes that can fit in the container. } 
    \label{tab:rubiks}
\end{figure*}

\subsection{Code Pattern Dimension}

Code decomposition at the introductory level is about identifying different tasks which are separated into different procedures that then work together in the solution program. This way of thinking about code decomposition is similar to problem decomposition which involves identifying the different tasks of the problem and how they relate~\cite{muller2007pattern}. In Pattern-Oriented Instruction (POI) pedagogy, problem decomposition is an intermediate step to solution creation starting from a problem prompt~\cite{muller2007pattern}. In the case of a code decomposition assignment, the student is provided with the problem prompt and solution, and then required to identify the individual tasks and how they relate for the purpose of decomposing them into separate procedures.

\begin{figure}[h]
    \centering
    \includegraphics[width=.9\columnwidth]{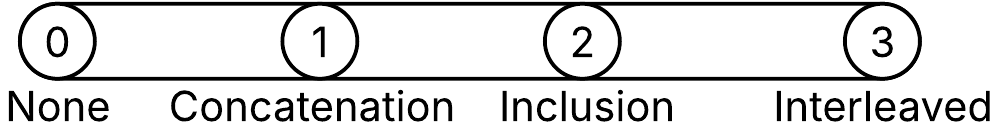}
    \caption{Levels of the Code Pattern Composition Dimension. Level 1 refers to code compositions where the tasks are completely separated, Level 2 refers to code compositions where one task is a inner part of a larger task and Level 3 refers to code compositions where the code switches from one pattern to another repeatedly. Code examples for these different types of code compositions are shown in Figure~\ref{tab:pattern_composition}.}
    \label{fig:code_pattern}
\end{figure}

We theorize that the complexity in separating the different tasks of a program comes from the way they relate to each other. Ginat et al.~\cite{ginat2013novice} have identified three ways in which tasks may relate to each other: concatenation, inclusion, and interleaved. Concatenation describes where two patterns are appended, one after the other. Inclusion describes when one pattern entirely appears inside another. Interleaving is used when pieces of two patterns are mixed. Ginat et al.\cite{ginat2009interleaved,ginat2013novice} has empirically shown that students have the most difficulty with the interleaved pattern composition. We rank concatenation at a lower level of complexity than inclusion: inclusion arises due to more complex nesting structure in code, as demonstrated by Figure \ref{tab:pattern_composition}. Figure ~\ref{fig:code_pattern} shows the different ways in which code patterns can interact in code in order of their complexity.

\begin{figure}[h]
    \centering
    \includegraphics[width=.9\columnwidth]{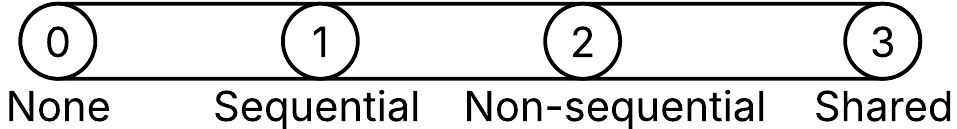}
    \caption{Levels of the Data Dependency Dimension.}
    \label{fig:data_dependency}
\end{figure}

\subsection{Data Dependency Dimension}

We reflect that the order of the statements that form a dependency path and whether different dependency paths have shared dependencies are the two main sources of complexity for the data dependency dimension, as shown in Figure~\ref{fig:data_dependency}. A shared dependency example is the Garden problem in Haldeman et al.~\cite{haldeman2025teaching}. The problem calculates the quantities of soil, fill and plants needed for a square garden with different geometric shapes; the formulas for each quantity need the area of an inner circle. The computation of each quantity represents a separate goal which can be separated out and implemented as its own function, and the computation of the inner circle is a shared dependency between them.

The non-sequential level is in between the sequential and shared levels because although the code statements with  shared dependency are not in the same order, there are no shared dependencies between them. The non-sequential placement of the code statements may obfuscate the components of each task. Thus, the student cannot rely on organizational cues in the code and must apply code comprehension skills to identify the individual tasks. The Rubik's Cube problem shown in Figure ~\ref{tab:rubiks} is an example of non-sequential data dependency. The problem has four tasks or goals: to compute the fit of a cube for all the three dimensions of a container which are then use to compute how many cubes can fit in the container by multiplying these three dimensions. Note that the code statements of the first three goals must be rearranged in the decomposed code.

\section{Using the proposed complexity framework}
\label{s:usage}

\begin{figure}[h]
    \centering
    \includegraphics[width=.5\textwidth]{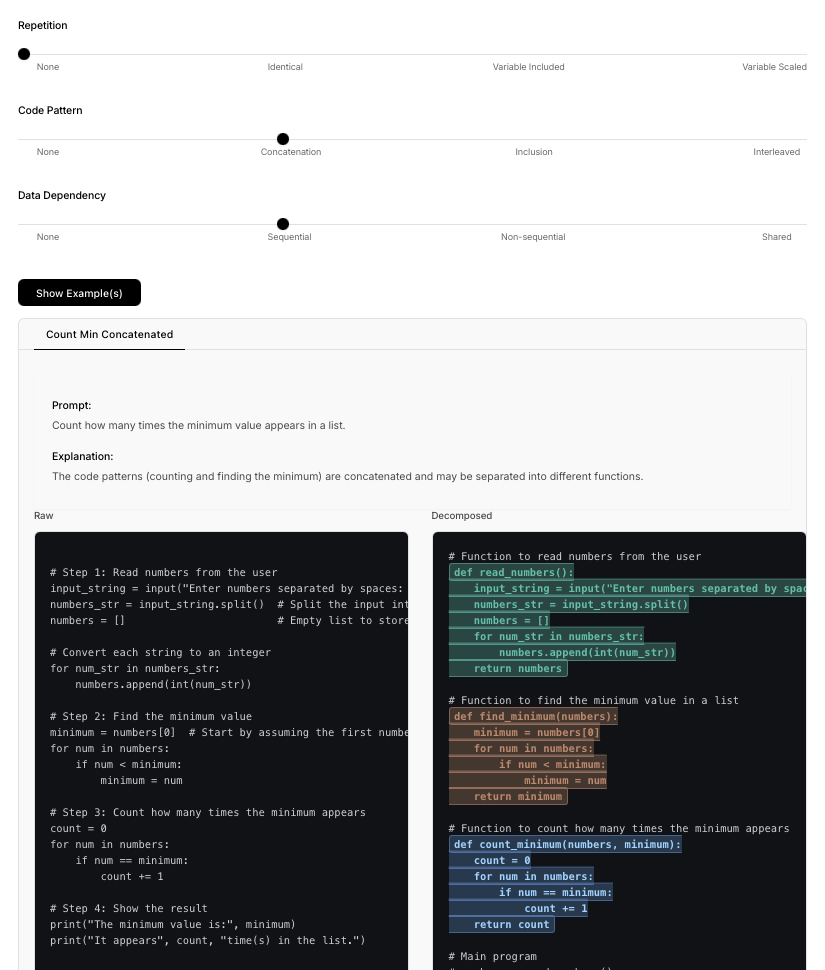}
    \caption{Example usage of our web tool. Note that users filter by data, code, and repetition pattern. Multiple examples may exist for a given category.}
    \label{fig:tool_eg1}
\end{figure}

As part of designing the framework, we collected problems from our prior work~\cite{haldeman2025teaching} and the sources listed in Table ~\ref{tab:sources}. Then we labeled them  using the proposed framework. Each label contains three tags, with each tag corresponding to the level description for one of the dimensions of the framework. We then used this information to create an online interactive repository of code structuring exercises\footnote{\url{https://georgianahaldeman.github.io/PDF/}} shown in Figure~\ref{fig:tool_eg1}. The repository has the following benefits over our previous educational resource~\cite{haldeman2025teaching}:

\begin{enumerate}
    \item It shows the starting unstructured code and the targeted decomposed code, the complexity level along the three dimensions, and additional information regarding the patterns and the structuring.
    \item It is searchable by complexity which facilitates the scaffolding of material. An instructor can select their desired levels of complexity along the three dimensions and then click \textit{Show Example(s)} which will retrieve one or more examples with the desired parameters.
    \item It can be easily extended with contributions from the community. Our hope is that instructors will employ the framework to first understand how to systematically think about code structuring exercises and then create new assignments that will be added to the repository.
\end{enumerate}

\section{Future directions}
\label{s:future}

This work provides an initial framework for thinking about the complexity of code structuring exercises, but there are several interesting avenues for future research and practice that can build upon it.

First, we would like to explore empirical validation of the framework with students. While the dimensions and levels proposed are grounded in prior literature, studies involving learners would allow us to evaluate whether the categories do indeed differentiate task difficulty. This could involve exploring how students perform across tasks with different combinations of repetition, code pattern composition, and data dependency levels. 

Another promising direction is the automated construction of code structuring tasks using generative AI.  Our existing repository (see Section \ref{s:usage}) was curated by the authors, but expanding this manually would take significant effort.  Generative AI models could be used to produce new tasks that are not only aligned with the framework dimensions, with automated tagging of complexity levels, but also personalized to match the interests of students.  Similar work involving the customization of learning resources has proven effective \cite{gutierrez2024evaluating, logacheva2024evaluating}.  On a related note, it may appear that as generative AI becomes increasingly powerful there will be a reduced need for humans to refactor or restructure code.  However, we argue that these skills remain essential, particularly at the introductory level. Even when AI systems generate code, students and future developers must be able to read, understand, and evaluate that code for correctness, efficiency, and maintainability. Practicing code structuring tasks builds critical code comprehension abilities that are essential for effective software development, even in an AI-assisted programming landscape.

Finally, it would be valuable to investigate how practicing bottom-up decomposition using such a resource impacts students' abilities in other areas, such as with top-down design, code refactoring, or algorithm development. Understanding such transfer effects would help to provide a strong case for integrating code structuring exercises into introductory courses.

\section{Conclusion}
\label{sec:conclusion}

In this 
paper we argue for increased attention to decomposition and abstraction in introductory computing courses; we further argue that repetition, pattern composition and data dependency are the main drivers of code structuring complexity. In addition, we propose a framework to systematically reason about the complexity of code structuring exercises at the introductory level. Finally, we provide an online inventory with a starting set of labeled exercises to serve the community in generating different combinations of scaffolded material. 
We invite the community to apply the framework in labeling existing problems, creating new ones, and contributing to the inventory.
Future work will focus on empirically validating the framework and using it to automatically generate a broad range of decomposition exercises.

\newpage

\bibliographystyle{ACM-Reference-Format}
\balance
\bibliography{sigcse2026,SE_books,references}

\end{document}